\documentclass[sigconf]{acmart}

\AtBeginDocument{%
  \providecommand\BibTeX{{%
    \normalfont B\kern-0.5em{\scshape i\kern-0.25em b}\kern-0.8em\TeX}}}

\setcopyright{acmcopyright}
\copyrightyear{2018}
\acmYear{2018}
\acmDOI{10.1145/1122445.1122456}

\acmConference[Woodstock '18]{Woodstock '18: ACM Symposium on Neural
  Gaze Detection}{June 03--05, 2018}{Woodstock, NY}
\acmBooktitle{Woodstock '18: ACM Symposium on Neural Gaze Detection,
  June 03--05, 2018, Woodstock, NY}
\acmPrice{15.00}
\acmISBN{978-1-4503-XXXX-X/18/06}

\usepackage{tikz}
\usepackage{amsmath}
\usepackage{amsmath,amsfonts}
\usepackage{algorithmic}
\usepackage{graphicx}
\usepackage{textcomp}
\usepackage{xcolor}
\usepackage{xspace}
\usepackage{comment}
\usepackage{multicol, multirow}
\usepackage{caption}
\usepackage{subfig}
\usepackage{booktabs}       
\usepackage[normalem]{ulem}

\begin{document}

\date{}

\newcommand{\zd}[1]{{\color{blue}{\it DZ: #1}}}
\newcommand{\dl}[1]{{\color{green}{\it DL: #1}}}
\newcommand{\xiang}[1]{{\color{purple}{[\it Xiang: #1]}}}
\newcommand{\sys}{DistDGLv2\xspace}

\title{Distributed Hybrid CPU and GPU training for Graph Neural Networks on
Billion-Scale Heterogeneous Graphs}

\author{Da Zheng}
\affiliation{%
  \institution{AWS AI}
  \country{USA}
}
\email{dzzhen@amazon.com}

\author{Xiang Song}
\affiliation{%
  \institution{AWS AI}
  \country{USA}
}
\email{xiangsx@amazon.com}

\author{Chengru Yang}
\affiliation{%
  \institution{AWS AI}
  \country{China}
}
\email{ychengru@amazon.com}

\author{Dominique LaSalle}
\affiliation{%
  \institution{NVIDIA Corporation}
  \country{USA}
}
\email{dlasalle@nvidia.com}

\author{George Karypis}
\affiliation{%
  \institution{AWS AI}
  \country{USA}
}
\email{gkarypis@amazon.com}

\begin{abstract}
Graph neural networks (GNN) have shown great success in learning from graph-structured data.
They are widely used in various applications, such as recommendation, fraud detection,
and search. In these domains, the graphs are typically large and heterogeneous,
containing many millions or billions of vertices and edges of different types.
To tackle this challenge, we develop \sys, a system that extends DistDGL for training
GNNs on massive heterogeneous graphs in a mini-batch fashion, using distributed
\textit{hybrid CPU/GPU} training. \sys places graph data in distributed CPU memory
and performs mini-batch computation in GPUs. For ease of use, \sys adopts API compatible
with Deep Graph Library (DGL)'s mini-batch training and heterogeneous graph API, which enables distributed
training with almost no code modification.
To ensure model accuracy, \sys follows a synchronous training approach and allows
ego-networks forming mini-batches to include non-local vertices.
To ensure data locality and load balancing,
\sys partitions heterogeneous graphs by using a multi-level partitioning algorithm
with min-edge cut and multiple balancing constraints.
\sys deploys an asynchronous mini-batch generation pipeline that makes computation and
data access asynchronous to fully utilize all hardware (CPU, GPU, network, PCIe).
The combination allows \sys to train high-quality models while
achieving high parallel efficiency and memory scalability.
We demonstrate \sys on various GNN workloads.
Our results show that \sys achieves $2-3\times$ speedup over DistDGL
and $18\times$ speedup over Euler. It takes only $5-10$ seconds to complete an epoch
on graphs with hundreds of millions of vertices on a cluster with 64 GPUs.

\end{abstract}


\maketitle

\section{Introduction}

Graph Neural Networks (GNNs) have shown success in learning from graph-structured data and have
been applied to many graph applications in social networks, recommendation,
knowledge graphs, etc. In these applications, graphs are usually huge and heterogeneous,
in the order of many millions or even billions of vertices and edges of different types.
For instance, Facebook's social network graph contains billions of users.
Amazon is selling billions of items and has billions of users and many sellers.
Natural language processing tasks take advantage of knowledge graphs,
such as Freebase \cite{freebase} with 1.9 billion triples and thousands of relations.

A number of GNN frameworks have been introduced that take advantage of distributed processing to scale GNN model training to large graphs. These frameworks differ on the type of training they perform (full-graph training vs mini-batch training) and on the type of computing cluster that they are optimized for (CPU-only vs hybrid CPU/GPU).
So far, few frameworks are designed to handle heterogeneous graphs with more than one vertex type and edge type.
Distributed frameworks that perform full-graph training have been developed for
both CPU- and GPU-based clusters~\cite{roc,neugraph,tripathy2020reducing,dorylus,distgnn}, whereas distributed frameworks that perform mini-batch training are mainly developed/optimized for CPU-based clusters~\cite{distdgl,euler,aligraph,agl}. Unfortunately, for large graphs, full-graph training is inferior to mini-batch training because it requires many epochs to converge and converges to a lower accuracy (cf., Sec~\ref{sec:minibatch}). This makes approaches based on distributed mini-batch training the only viable solution for large graphs. However, before such mini-batch-based approaches can fully realize their potential in training GNN models for large graphs, they need to be extended to take advantage of GPUs' higher computational capabilities. 

%

It is natural to ask whether GPUs have advantage of training GNN models on large graphs.
The main challenges of GNN training on GPUs lie in two aspects. First, GNN models have much
lower computation density than traditional neural network models, such as CNNs and
Transformers. Consequently, for very large graphs, since we cannot store the entire graph 
and all of its features in GPU memory, it is critical to devise efficient strategies 
for moving data from slower memory (e.g., CPU, remote memory, disks) to GPUs during training.
%
%
The second challenge is load imbalance among mini-batches. Typically, neural network 
models are trained with synchronous stochastic gradient descent (SGD) to achieve 
good model accuracy, which requires a synchronization barrier at the end of every 
mini-batch iteration. To ensure good load balance, mini-batches have to contain the 
same number of vertices and edges as well as reading the same amount of data from slower
memory. Due to the complex subgraph structures in natural graphs, it is difficult to
generate such balanced mini-batches. The load balancing problem becomes even more
severe on heterogeneous graphs because vertices of different types may be associated with
different feature sizes.


In this work, we develop \sys on top of DGL~\cite{dgl} to optimize distributed GNN
training on heterogeneous graphs for hybrid CPU/GPU clusters, where it stores
the graph structure and vertex/edge features in CPU memory and performs mini-batch
computation in GPUs.
To provide good user experience and to minimize accuracy differences between development
and deployment of a model, \sys provides Python API compatible with
DGL's mini-batch sampling and heterogeneous graph API. Thus, it requires almost
no code modification to DGL's training
scripts to enable distributed training.
To ensure the quality of GNN models, \sys uses synchronized SGD and generates mini-batches
with non-local vertices.
It extends the design principles of DistDGL~\cite{distdgl}, a CPU-only distributed
GNN training framework, to increase data locality and balance computation among trainers on
heterogeneous graphs. It deploys a sophisticated asynchronous mini-batch
sampling pipeline that performs computation and data access ahead of time only
on immutable data to overlap CPU and GPU computation and data communication and
utilize all hardware resources (CPU, GPU, network, PCIe) simultaneously. Thus, \sys
accelerates mini-batch training without changing the mini-batch sampling algorithm.

We conduct comprehensive experiments to evaluate the efficiency of \sys.
Overall, \sys achieves $18\times$ speedup
over Euler-GPU and $2-3\times$ speedup over DistDGL-GPU,
the modified version of Euler \cite{euler} and DistDGL~\cite{distdgl} for GPU training,
on a cluster of 32 GPUs.
\sys achieves up to $15\times$ speedup over distributed CPU
training by DistDGL in a cluster of the same size. This indicates that GPUs can be
effective for GNN mini-batch training on massive graphs than CPUs.
It takes 5 seconds per epoch to train GraphSage and GAT on
a homogeneous graph
with 100 million vertices and 13 seconds per epoch to train RGCN on a heterogeneous graph
with 240 million vertices with 64 GPUs.

The main contributions of the work are listed below:
\begin{itemize}
    \item We design an asynchronous mini-batch sampling pipeline with extensible Python sampling API
    and speed up distributed GNN training on
    hybrid CPU/GPU by a factor of $2-3\times$ over DistDGL-GPU without changing the training algorithm.
    \item \sys is a distributed GNN framework that explicitly supports
    distributed heterogeneous graphs with very diverse vertex/edge features.
    \item \sys realizes all optimizations under DGL's API for ease of use
    and minimizing accuracy differences between development and deployment of a model.
\end{itemize}
\section{Related Work}


Many works have been developed to scale GNN training on large graph data for distributed CPU- and GPU-based clusters.
Many of them~\cite{roc,neugraph,tripathy2020reducing,dorylus,distgnn} are designed for distributed full-graph training on multiple GPUs or distributed memory whose aggregated memory fit the graph data.
Even though full-graph training is easier to parallelize, it actually takes
a longer time to converge on a large graph and may converge to a lower accuracy
than mini-batch training (Section \ref{sec:minibatch}). Therefore, the focus of
our work is to optimize mini-batch training.

\begin{figure}
\centering
\includegraphics[scale=0.5]{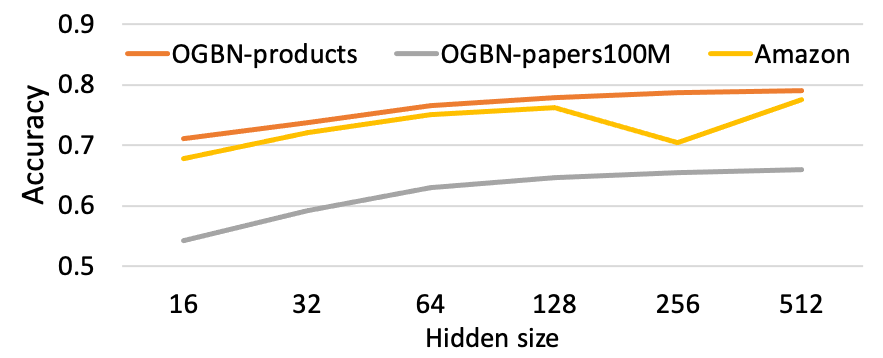}
\vspace{-3mm}
\caption{The model accuracy of GraphSage with different hidden sizes on datasets in
Section \ref{sec:eval}.}
\label{fig:hidden_size_vs_acc}
\end{figure}

Multiple frameworks have been developed for distributed GNN mini-batch training.
Some of them \cite{aligraph, agl, euler} adopt distributed mini-batch training
but does not use graph partitioning algorithms that minimize edge cut to
reduce network communication.
Their system is optimized for distributed training on a CPU cluster and
many of their design choices (e.g., only using multiprocessing) are not
suitable for GPU training.
DistDGL~\cite{distdgl} adopts METIS graph partitioning~\cite{metis} to
reduce network communication, but is not designed for GPU training.
Frameworks, such as PyTorch-Geometric~\cite{pyg} and PaGraph \cite{pagraph},
support multi-GPU training but cannot scale to graphs beyond the memory capacity of
a single machine.
P3~\cite{p3} is a distributed GNN framework designed for distributed training in a GPU cluster.
It adopts model parallelism, which works better when the hidden size is small.
In contrast, \sys uses data parallel and works better for larger hidden sizes.
As shown in Figure \ref{fig:hidden_size_vs_acc},
a large hidden size is required to achieve good model accuracy.
BGL~\cite{bgl} builds on top of DGL for distributed training. It heavily relies on
changing the mini-batch sampling algorithm to increase data locality and GPU cache hits.
This is orthogonal to \sys's design.
\sys focuses on optimizations that are agnostic to models and training
algorithms to provide robust model training. It offers
very flexible mini-batch sampling pipeline to adopt more advanced sampling algorithms.
\section{Background}

\subsection{Graph neural networks} \label{sec:gnn}

GNNs emerge as a family of neural networks capable of learning a joint representation
from both the graph structure and vertex/edge features.
Recent studies~\cite{gnn-chemistry, graphnets} formulate GNN models with
\emph{message passing}, in which vertices broadcast messages to their neighbors and compute
their own representation by aggregating messages.

More formally, given a graph $\mathcal{G(\mathcal{V},\mathcal{E})}$, we denote
the input feature of vertex $v$ as $\mathbf{h}_v^{(0)}$, and the feature of the edge
between vertex $u$ and $v$ as $\mathbf{e}_{uv}$.
To get the representation of a vertex at layer $l$, a GNN model performs the computations
below:

\vspace{-0.5em}

\begin{equation}\label{eq:mp-vertex}
\mathbf{h}_v^{(l+1)} = g(\mathbf{h}_v^{(l)},\bigoplus_{u\in\mathcal{N}(v)} f(\mathbf{h}_u^{(l)}, \mathbf{h}_v^{(l)}, \mathbf{e}_{uv}))
\end{equation}

\vspace{-0.5em}

\noindent $f$, $\bigoplus$ and $g$ are customizable or parameterized functions
for generating messages, aggregating messages and
updating vertex representations, respectively.
Similar to convolutional neural networks (CNNs), a GNN model iteratively applies
Equations~\eqref{eq:mp-vertex} to generate representations with multiple layers.


\subsection{Mini-batch training} \label{sec:minibatch}
\begin{figure}[h]
\centering
\includegraphics[scale=0.45]{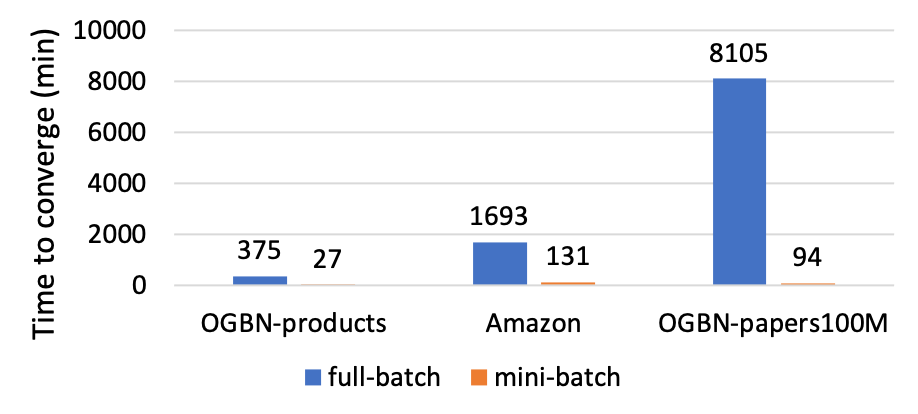}
\vspace{-3mm}
\caption{Train GraphSage with full-graph and mini-batch training on medium-size and large graphs on the same CPU machine using DGL.}
\label{fig:full_batch_vs_mini_batch}
\end{figure}

Even though GNN models can be trained in full-batch fashion, mini-batch training
is more practical for GNN models on large graphs.
It has been established that training neural networks with SGD using small mini-batches
converges faster and to a lower minimal than passing the whole dataset through the network
\cite{wilson2003general, lecun2012efficient, keskar2016large}.
Figure \ref{fig:full_batch_vs_mini_batch} shows the time of full-graph and mini-batch
training to converge on graphs of medium scale and large scale (Table \ref{tbl:dataset})
on the same CPU machine. On these graphs, full-batch training of GraphSage
is one or two orders of magnitude slower than mini-batch training. In addition,
full-graph training cannot converge to the same accuracy as mini-batch training on some graphs.
For example, full-graph training on the Amazon dataset has the test accuracy
of 0.68 while mini-batch training gets test accuracy of 0.77.

GNN mini-batch training is different from other neural networks due to
the data dependency between vertices. We need to carefully sample subgraphs to capture
the dependencies in the original graph.
A typical strategy of mini-batch sampling for GNN \cite{graphsage} follows three steps:
(i) sample a set of $N$ vertices, called \textit{target vertices}, uniformly at random from
the training set; (ii) randomly pick
at most $K$ (called \textit{fanout}) neighbor vertices for each target vertex.
When the GNN has multiple layers, neighbor sampling repeats recursively.
That is, from a sampled neighbor vertex, it continues sampling
its neighbors. The number of recursions is determined by the number of layers
in a GNN model. This sampling algorithm results in mini-batches with 100s times more
vertices than the number of target vertices and causes large amount of
communication in distributed training.



\subsection{Distributed training}
\begin{figure}
\centering
\includegraphics[width=0.9\linewidth]{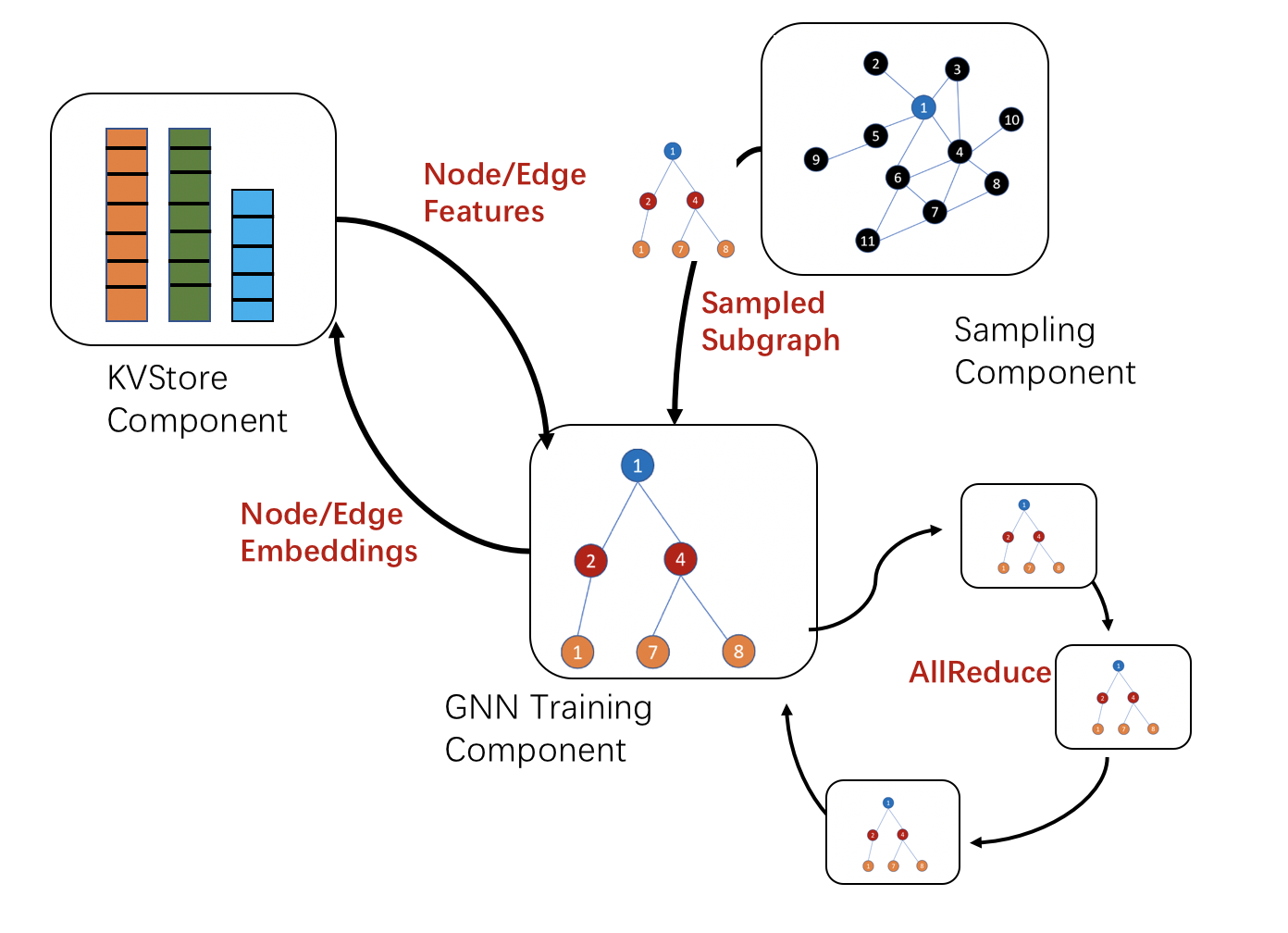}
\vspace{-3mm}
\caption{Distributed training components in DistDGL.}
\label{fig:dist}
\end{figure}

DGL/DistDGL \cite{distdgl} provides the distributed training capability
on homogeneous graphs. It uses the existing programming interface of mini-batch
training in DGL and divides distributed training into three components
(Figure \ref{fig:dist}):
\begin{itemize}
    \item A \textit{mini-batch sampler} samples mini-batches from the input graph.
    Users invoke DistDGL samplers in the trainer process. Internally, the sampling
    requests are handled by multiple sampling processes, which generates
    remote process calls (RPC) to perform distributed sampling.
    \item A \textit{KVStore} that stores all vertex features and edge features across machines.
    It provides the \textit{pull} and \textit{push} interfaces for pulling data from
    or pushing data to the distributed store.
    \item A \textit{trainer} fetches mini-batch graphs from the sampler and
    corresponding vertex/edge features from the KVStore and runs the forward
    and backward computation to compute the gradients of the model parameters.
\end{itemize}

When DistDGL deploys these logical components to actual hardware, it is mainly
optimized for distributed training in CPU, in which the main optimization
is to reduce the network traffic among machines. DistDGL partitions the input graph
with METIS algorithm \cite{metis} and partitions the vertex/edge features according to
graph partitions. DistDGL launches sampler servers, KVStore servers
and trainers on the same cluster of machines and dispatches computation to the data
owner to reduce network communication.

\section{System Design}
\sys preserves the programming interface of DGL/DistDGL and extends DistDGL
in two major ways. First,
It implements distributed heterogeneous graphs with guarantees in load balancing
and data locality and expose DGL's heterogeneous graph
interface for ease of use.
It optimizes distributed hybrid CPU/GPU training, where graph data are in
distributed CPU memory and mini-batch computation in GPU, with
API compatible to DGL’s mini-batch training. As such, \sys enables
distributed training with almost no code modification to DGL’s training scripts.
To construct an efficient system for distributed hybrid CPU/GPU training,
we optimize the system in three aspects.


\textit{Data locality}:
Reducing data movement from slower memory (e.g., remote memory) to GPUs is essential
to the training speed.
To reduce data movement from the distributed CPU memory to GPUs,
\sys partitions a heterogeneous graph with the METIS algorithm and co-locates
trainers with graph partitions
(Section \ref{sec:partition}). To minimize data copy in CPU inside a trainer,
it only uses multithreading for parallelization in the mini-batch sampling pipeline
(Section \ref{sec:async_pipeline}). To reduce data copy to GPUs in a mini-batch,
\sys deploys two-level graph partitioning to reduce the number of vertices in a mini-batch
(Section \ref{sec:two_level_partition}).

\textit{Load balancing}:
Due to neighbor sampling, GNN mini-batches may vary significantly in the number of
vertices and edges. In a heterogeneous graph, vertices of different types may have
different features, which makes data access more imbalanced if vertex features
are not evenly distributed among machines.
\sys balances the distributed training workloads in two levels.
In the data preprocessing, it ensures roughly the same number
of vertices and edges of different types in each partition
(Section \ref{sec:part_balance}).
During training, it removes global synchronization barrier in mini-batch generation
to hide the impact of any imbalance in mini-batch sampling from the training process
(Section \ref{sec:async_pipeline}).

\textit{Use all hardware resources simultaneously}:
Distributed hybrid CPU/GPU training involves in different hardware resources:
CPU, GPU, network, PCIe, etc. Different hardware has different computation speeds
or data transfer speeds. To use all hardware resources effectively,
\sys adopts two separate strategies: 1) split mini-batch generation into many stages in
a pipeline and turn all computations into asynchronous operations
(Section \ref{sec:async_pipeline}) to overlap computation and communication,
2) move more computation to GPUs to reduce
the burden in CPU (Section \ref{sec:dist_sample}).

\subsection{Distributing heterogeneous graphs}
\begin{figure*}
\centering
\includegraphics[width=0.9\linewidth]{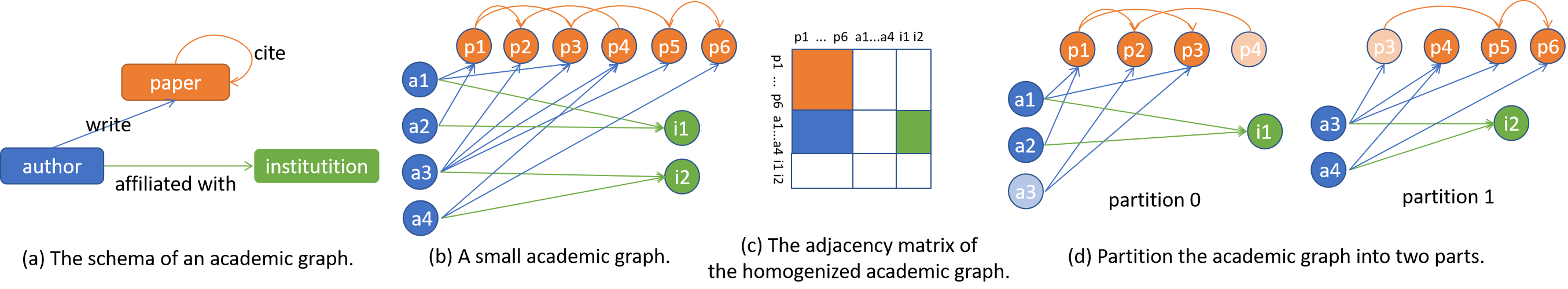}
\vspace{-3mm}
\caption{An example of a heterogeneous graph and its distributed storage.}
\label{fig:heterograph}
\vspace{-1em}
\end{figure*}

\sys is designed to support heterogeneous graphs with diverse vertex and edge features
while providing the same user-friendly heterogeneous graph API of DGL.
Figure \ref{fig:heterograph} (b) shows a heterogeneous graph whose schema is
shown in Figure \ref{fig:heterograph} (a). After partitioning a heterogeneous
graph and storing data in a cluster of machines (Figure \ref{fig:heterograph} (d)),
\sys's API allows users to access data in the distributed graph as if accessing a graph
in a single machine. When partitioning a heterogeneous graph, we ensure minimal edge cut
and balanced partitions.

\subsubsection{Partition heterogeneous graphs} \label{sec:partition}
To reduce data communication in distributed training, \sys deploys METIS \cite{metis}
to partition a heterogeneous graph with a minimal number
of edge cuts across partitions. Because graph partitioning is a preprocessing step,
the partitioning overhead can be amortized. We usually partition a graph
once and use it for many training runs (e.g., during hyperparameter tuning).


Because METIS can only partition a homogeneous graph, \sys homogenizes a heterogeneous
graph and stores the entire graph
in a single adjacency matrix, in which all vertices, regardless of their vertex types,
are assigned with unique vertex IDs (Figure \ref{fig:heterograph} (c)). The edges are
located in the colored blocks in the adjacency matrix. In this format,
the vertices and edges of the same type are assigned with contiguous IDs and
vertex types and edge types are stored as metadata.
We pass the adjacency matrix to METIS for partitioning, which results in
partitions in Figure \ref{fig:heterograph} (d).
After assigning vertices to a partition, \sys follows the same strategy in DistDGL
to assign edges to partitions and split a graph into physical subgraphs.
This results in partitions shown in Figure \ref{fig:heterograph} (d).

\subsubsection{Hierarchical partitioning} \label{sec:two_level_partition}
\sys deploys two-level partitioning to reduce data transfer to GPUs.
In the first level, we deploy METIS to split a graph into physical subgraphs
and assign one first-level partition to a machine. Due to the min-edge cut
by METIS, the first-level partitions reduce data communication across the network.
Inside each partition, we run METIS again to generate second-level partitions
and assign one second-level partition to a GPU.
Instead of generating physical subgraphs for the second level, we simply
assign vertices to second-level partitions and split the training set accordingly.
As such, a trainer samples target vertices or edges
from the local second-level partition. This increases locality in neighbor sampling.
That is, two vertices is more likely to sample the same neighbor vertex,
which reduces the number of vertices in a mini-batch. Our ablation study
(Section \ref{sec:ablation}) shows that introducing the second-level partitions
can effectively reduce the number of vertices in a mini-batch and improves
the training speed by roughly 20\% on the benchmark datasets.



\noindent Because \sys uses synchronous SGD to train
the model, the estimation of the model gradients is unbiased. As such,
distributed training in \sys in theory does not affect the convergence rate
or the model accuracy.

\subsubsection{Load balancing on graph partitions} \label{sec:part_balance}
Minimizing edge cut reduces data communication in distributed training,
but may
result in imbalanced partitions and imbalanced data storage in the cluster.
In a heterogeneous graph, different vertex types and edge types may be associated with
different data sizes. It is essential to distribute graph partitions, vertex data and
edge data of different types evenly across all machines
so that CPU memory storage in each machine is fully utilized and data access
during the training can be evenly distributed among machines.
By default, METIS only balances the number of
vertices in a graph. This is insufficient for a heterogeneous graph. 
We formulate this load balancing
problem as a multi-constraint partitioning problem~\cite{multiconstraint}.
\sys takes advantage of the multi-constraint mechanism in METIS to
balance training/validation/test vertices/edges in
each partition as well as balancing the vertices of different types and
the edges incident to the vertices.

\subsubsection{Heterogeneous graph vertex/edge data}
To support flexible storage of diverse features on different vertex types and edge types,
\sys extends the distributed KVStore to store features on each vertex type and edge type
separately.
The extended KVStore supports an arbitrary number of ID spaces. When
\sys loads a distributed heterogeneous graph, it creates an ID space in KVStore
for each vertex type and edge type. Each ID space is also associated with
a partition policy that maps vertex/edge data to physical machines.
The partition policy is derived from the first-level graph partitioning
(Section \ref{sec:partition}).
To simplify the access to vertex/edge data in the KVStore, \sys uses type-specific
vertex/edge ID, which is only unique within a particular vertex/edge type.

\subsection{Asynchronous mini-batch generation}

The key of efficient hybrid CPU/GPU training is to bring mini-batch data to GPU
efficiently. This requires optimizations in multiple aspects. First, we need to
overlap mini-batch generation with
mini-batch computation as well as overlapping computation and communication
to simultaneously utilize all computation resources (e.g., CPU and GPU) and
communication channels (e.g., network, CPU memory and PCIe). We need to
parallelize computation in CPU and avoid any unnecessary data copy in CPU.
In addition, we need to hide the impact of imbalance of GNN mini-batch sampling among
different trainers.

\subsubsection{Asynchronous mini-batch pipeline} \label{sec:async_pipeline}
\begin{figure*}%
    \centering
    \subfloat[\centering The stages of the mini-batch pipeline. The arrows indicate the dependencies of computation.]{{\includegraphics[width=14cm]{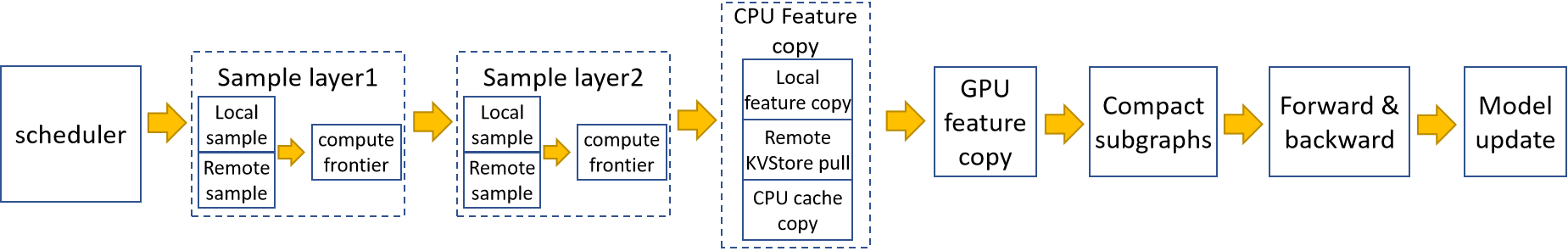} }}%
    \qquad
    \subfloat[\centering Realize the asynchronized mini-batch pipeline with three threads.]{{\includegraphics[width=14cm]{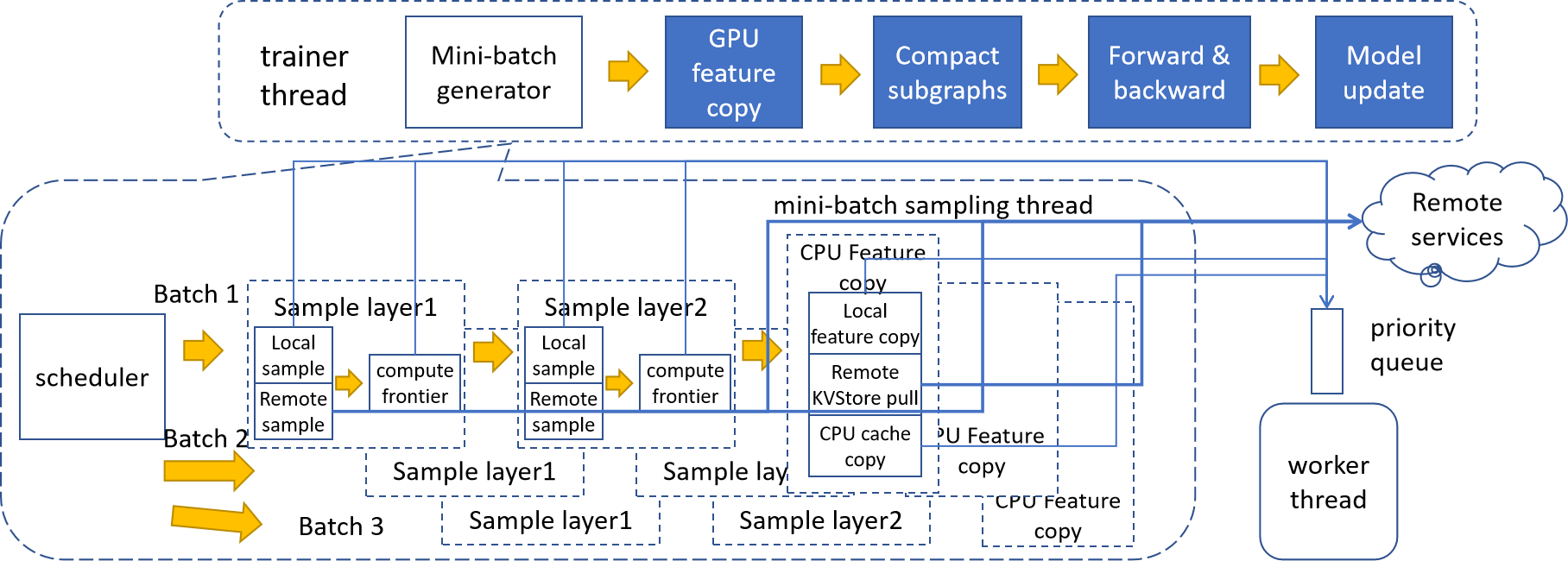} }}%
    \vspace{-3mm}
    \caption{\sys deploys an asynchronous mini-batch pipeline for hybrid CPU/GPU training.
The pipeline are divided into multiple stages. Some of the stages run in GPUs, indicated
by the blue boxes, while others run in CPU, indicated by the white boxes.
The computations in the pipeline run in three threads. All GPU computations
are invoked in the trainer thread; sampling computation and CPU feature copy
are invoked in the sampling thread but their actual computation happens in
the worker thread. In the sampling thread, computations in multiple mini-batches
are invoked simultaneously in a pipelining fashion to overlap computation of
different stages.}%
    \label{fig:pipeline}%
\end{figure*}

\sys deploys an asynchronous pipeline that generates mini-batches from
the distributed graph. It provides the sampling API compatible with DGL and delivers
mini-batches from a distributed graph as if sampled from a graph in a single machine.
It allows customization
of sampling algorithm in Python while deploying heavy optimizations to speed up
computation.
\sys divides mini-batch training into many stages (Figure \ref{fig:pipeline} (a)):
\vspace{-0.2em}
\begin{itemize}
\item a scheduler that
determines target vertices or target edges in each mini-batch to support various
learning tasks (e.g., node classification, link prediction) for GNN models,
\item neighbor sampling that samples multi-hop neighbors of the target vertices for
GNN computation,
\item CPU feature copy that fetches data from both local machines and remote machines
for each mini-batch and stores data in contiguous CPU memory,
\item GPU feature copy that loads data from CPU to GPU,
\item post-sampling GPU processing for mini-batches (in vertex-wise neighbor sampling,
we performs subgraph compaction that remaps vertex IDs and edge IDs in the subgraph
in GPU),
\item forward and backward computation on mini-batches,
\item model parameter updates.
\end{itemize}

There are dependencies between operations in different stages, but in some stages
there are multiple operations that can run in parallel.
For example, the stage of \textit{CPU feature copy} requires the frontier of
the input layer to be complete in the sampling stage;
on the other hand, copying features in CPU includes data copy from the local partition,
from remote KVStore and from local CPU cache, which can run independently.
A neighbor sampling stage can be further divided into two substages: sample neighbors
and compute the frontier. The two substages also have dependencies: we have to wait
for neighbor sampling to complete before computing the frontier of the layer.

\sys implements a flexible and efficient asynchronous mini-batch pipeline
to overlap computation with network communication
(Figure \ref{fig:pipeline} (b)). Because the target vertices or edges are
sampled from the training set randomly, there are no dependencies between mini-batches.
This allows us to sample mini-batches ahead of time and process multiple mini-batches
in a pipelining fashion. \sys divides the mini-batch pipeline into two parts:
1) mini-batch sampling in CPU, which includes mini-batch scheduling, distributed neighbor
sampling and CPU feature copy, and 2) post-sampling computation in GPU, which includes
data loading to GPUs, compacting subgraphs, 
and mini-batch computation in GPU. To avoid sampling computation from blocking
mini-batch training in GPU, \sys creates a dedicated Python thread for
the sampling computation in CPU, which allows us to run customized Python code for
sampling. We refer to this thread as a \textit{sampling thread}
and the original thread as a \textit{training thread}.
The CPU and GPU division reduces the interference between the two threads:
The GPU computation in the training thread has a global synchronization barrier
among trainers due to synchronized SGD but is not blocked by any computation
in the sampling thread; the sampling thread is not blocked by the global
synchronization barrier caused by SGD. Even though Python threads have a global lock to guard
the data access to Python objects, the lock is released whenever we jump to C code.
Thus, the Python global lock does not interfere the computation of the two threads
by much.

Inside the sampling thread, \sys performs all computation asynchronously
to ensure that network operations are not blocked by local CPU computations.
Whenever \sys performs a local operation in CPU, it creates a job for this operation,
sends it to a worker thread and returns immediately. The job is placed in a priority
queue of the worker thread. A job created for a later stage gets a higher priority.
Whenever \sys performs a remote operation, it issues an RPC request and returns
immediately. The sampling thread processes multiple mini-batches simultaneously.
After it processes the operations of a mini-batch at a certain stage, instead of waiting
for the operations to be complete, it proceeds to some operations of another mini-batch
at a different stage. After issuing a sufficient number of pending operations,
the thread sleeps and waits for some operations to be complete.

This aggressive ahead-of-time mini-batch generation can potentially lead to data
staleness and consume much memory. Currently, we only apply this ahead-of-time
mini-batch generation on immutable data (i.e., sampling from the graph structure
and read vertex/edge features). If a model has learnable embedding table on vertices,
we read the learnable embeddings synchronously. Thus, our asynchronous mini-batch
pipeline does not affect model convergence at all. To reduce memory consumption,
the pipeline sets different capacities for different stages. The capacity is defined
with the number of pending operations issued in a stage. The memory consumption
by the operations at different stages is different.
At the beginning of the pipeline (mini-batch scheduling and neighbor sampling),
a pending operation only needs to store
vertex IDs and edge IDs, which does not require too much memory, so we can work on
many mini-batches simultaneously. In the middle of the pipeline, we prefetch
vertex/edge features from remote machines and collect features
from local partitions, which may require hundreds of megabytes of CPU memory,
so we allows a relatively small number of mini-batches.
At the end of the pipeline, we only move one mini-batch ahead of time to GPUs
because of the scarceness of GPU memory. As such, we use a relatively large capacity
for scheduling and neighbor sampling (e.g., $25$); a relatively small
capacity (e.g., $5$) for CPU feature copy; a capacity of $1$
for GPU feature copy.

The main reason of using multithreading for parallelizing sampling in \sys is to
minimize data copy in the pipeline.
This is different from many other distributed GNN training frameworks, such as
DistDGL~\cite{distdgl} and Euler~\cite{euler}, which uses multiprocessing
for parallelization.
It is essential to minimize data copy in data intensive workloads such as graph
computation. Even though
multiprocessing parallelizes sampling computation well, it requires
to copy mini-batch data between processes, which results in additional
data copy and data serialization and deserialization. In contrast,
multithreading completely avoids these overheads. To further reduce data copy
in the pipeline, \sys carefully manage data buffers for network
communication and data copy between CPUs and GPUs. It allocates a pinned memory
buffer to collect data from the network and from the local partition before
sending them to GPUs, which results in only one data copy for each byte.

The asynchronous sampling pipeline introduces a startup overhead when filling
the pipeline at the beginning of every epoch. This hurts
the performance especially when the training set is small.
To remove the startup overhead, we run the asynchronous sampling pipeline throughout
the entire training without stopping in the sampling thread. The trainer thread only needs
to fetch mini-batches from the sampling thread.


\subsubsection{Distributed hybrid CPU/GPU sampling} \label{sec:dist_sample}
In hybrid CPU/GPU training, the distributed graph is placed in CPU memory. We have to
sample neighbors of target vertices
on CPU from the distributed graph. To take advantage of GPU's computation power,
we move some computation to GPUs. As such,
\sys divides the computation into multiple components. In this section,
we use the vertex-wise neighbor sampling algorithm \cite{graphsage}
for vertex classification for illustration. The same computation decomposition
applies to other neighbor sampling algorithms, such as layer-wise sampling
\cite{ladies}, and to other training tasks, such as link prediction.

Figure \ref{fig:pipeline} shows the mini-batch sampling pipeline of vertex-wise
neighbor sampling \cite{graphsage} for vertex classification. It starts from
the seed vertices and samples their neighbor vertices in the ego-network.
The sampling computation proceeds with one hop of neighborhood at a time.
After sampling all neighbors within a hop, it computes the frontier (i.e.,
the unique set of vertices) as the seed vertices for the next-hop
neighbor sampling.

\sys divides neighbor sampling into two components.
In CPU, it samples vertices and edges
from the distributed graph for each hop of neighborhood. The sampled subgraphs
are small enough to fit in GPU memory. It then moves the subgraphs to GPU and
performs graph compaction to remove empty vertices and relabel vertices and edges
for mini-batch computation.
This algorithm samples neighbors on each vertex independently and, thus, we can
further decompose the sampling computation within a hop into local sampling and
remote sampling. \sys dispatches the remote sampling requests to the sampler servers
and issues a job for local sampling simultaneously.
After local and remote sampling complete, \sys collects the results from different
partitions, stitches them together and issues another job to compute
the frontier vertices for the next hop.
All sampling computation within a hop runs in CPU.
To better parallelize the frontier computation, \sys bundles
the sampling computation of multiple mini-batches and use OpenMP to parallelize them.


\section{Evaluation}\label{sec:eval}
In this section, we evaluate \sys with multiple GNN models on large graph datasets.
We benchmark three commonly used GNN models (GraphSAGE~\cite{graphsage},
Graph Attention Networks (GAT)~\cite{gat} and Relational Graph Convolution Networks
(RGCN)~\cite{rgcn}) to evaluate the performance of \sys.

\begin{table}
\footnotesize
    \centering
    \caption{Dataset statistics.}\label{tbl:dataset}
    \begin{tabular}{@{\hspace{1pt}}l@{\hspace{8pt}}r@{\hspace{8pt}}r@{\hspace{8pt}}r@{\hspace{8pt}}r@{\hspace{8pt}}r@{\hspace{1pt}}}\toprule
         Dataset & \# Vertices & \# Edges & Vertex & \# train & \# train \\
                 &          &             & features & vertices & links \\
         \midrule
         \textsc{ogbn-product}\cite{ogb} & 2.4M & 61.9M & 100 & 197K & 61.9M \\
         \textsc{Amazon} \cite{clustergcn} & 1.6M & 264M & 200 & 1.3M & 264M \\
         \textsc{ogbn-papers100M}\cite{ogb} & 111M &  3.2B & 128 & 1.2M & 3.2B \\
         \textsc{ogbn-mag}\cite{ogb} & 1.9M & 21M & 128 & 629K & 21M \\
         \textsc{mag-lsc}\cite{ogb-lsc} & 240M & 7B & 756 & 1.1M & 7B \\
         \bottomrule
    \end{tabular}
\end{table}

Our benchmarks use three medium-size graphs (\textsc{ogbn-product} \cite{ogb}, \textsc{Amazon}
\cite{clustergcn} and \textsc{ogbn-mag}) and two large
graphs (\textsc{ogbn-papers100M} \cite{ogb} and \textsc{mag-lsc} \cite{ogb-lsc}) (see~Table~\ref{tbl:dataset} for various statistics). Note that even though all datasets contain labels for vertex classification, the number of labeled vertices in all but the smaller datasets is similar. As a result, the cost to train vertex classification models for the larger graphs is not as high as the size of the graphs suggests. 
However this is not the case for the link-prediction task, for which we use all the edges to train the GNN models, leading to training sets with billions of data points.

\subsection{\sys vs. other distributed GNN frameworks}

We compare the training speed of \sys with DistDGL~\cite{distdgl} and Euler~\cite{euler},
the state-of-the-art distributed GNN mini-batch training frameworks, on four g4dn.metal
instances, each of which is equipped with eight NVIDIA T4 GPUs and two Intel Xeon Platinum 8259CL CPUs, for a total of 32 GPUs and 192 CPU cores. Both DistDGL and Euler are designed for distributed CPU training, so we
run them on four r5dn.24xlarge instances to collect their CPU training speed, each of which have two Intel Xeon Platinum 8259CL CPUs, for a total of 192 CPU cores across the four instances.
To have a fair comparison with \sys, we change DistDGL and Euler to perform GNN training
on GPUs
by moving sampled mini-batches to GPUs. We refer to their CPU versions as DistDGL-CPU
and Euler-CPU and their GPU versions as DistDGL-GPU and Euler-GPU.
We run all experiments with the same global batch size (the total size of
the batches of all trainers in an iteration) to get the same convergence.

\begin{figure}%
    \centering
    \subfloat[\centering GraphSage and GAT on homogeneous graphs]{{\includegraphics[width=4cm]{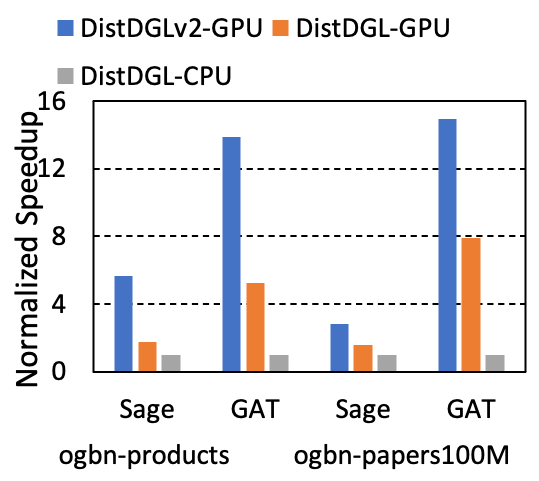} }}%
    \qquad
    \subfloat[\centering RGCN on heterogeneous graphs]{{\includegraphics[width=3.5cm]{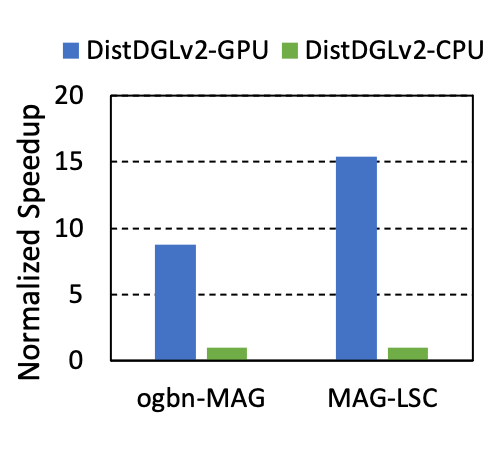} }}%
    \vspace{-3mm}
    \caption{Training speed of \sys compared with DistDGL.}%
    \label{fig:vs_dgl}%
\end{figure}

Figure \ref{fig:vs_dgl} (a) shows that \sys gets $2-3\times$ speedup over DistDGL-GPU
on various datasets.
\sys has higher speedup over DistDGL-GPU
on simpler GNN models (e.g., GraphSage). The main bottleneck of GraphSage training
is mini-batch sampling in CPU and data copy to GPUs. Even though
both \sys and DistDGL use METIS to partiton a graph and co-locate data with computation,
this alone cannot fully take advantage of GPU's computation. Asynchronous mini-batch
generation, parallelization strategies and load balancing deployed in \sys further
improve the performance of GNN training.

To verify the benefit of distributed GNN training with GPUs, we compare \sys-GPU with
DistDGL-CPU on GraphSage and GAT and \sys-CPU on RGCN.
Figure \ref{fig:vs_dgl} shows \sys-GPU has up to $15\times$ speedup over DistDGL-CPU and
\sys-CPU, which indicates that high
floating-point computation and fast memory in GPU are beneficial to train GNN models
on large graphs especially for more complex GNN models, such as GAT and RGCN.
Even for GraphSage, using GPUs still gets $3-6\times$ speedup.

\begin{figure}[h]
\centering
\includegraphics[scale=0.6]{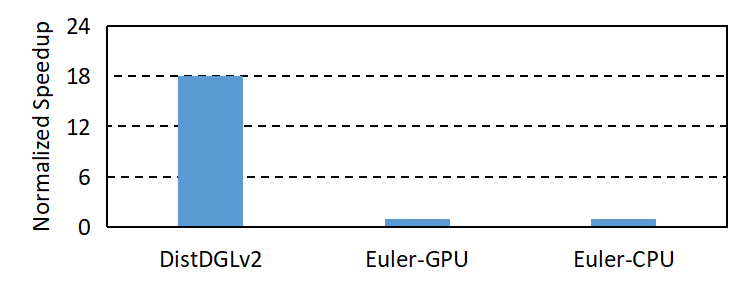}
\vspace{-3mm}
\caption{The speedup of \sys and Euler-GPU over Euler-CPU
for training GraphSage on \textsc{ogbn-product}.}
\label{fig:vs_euler}
\end{figure}

We further compare \sys with Euler on CPUs and GPUs when training GraphSage on
\textsc{ogbn-product} (Figure \ref{fig:vs_euler}). \sys gets $18\times$ speedup over
both Euler-CPU and Euler-GPU. Euler-GPU does not get speedup over Euler-CPU. Because
Euler only uses multiprocessing to parallelize computation and run sampling inside
the trainer process, it requires many trainer processes to achieve good performance.
This parallelization strategy works relatively well on CPU clusters but
does not work well on GPUs because we usually launch one trainer process
per GPU to save GPU memory and avoid interfering computation between trainer processes
on the same GPU. This indicates that effective distributed
GNN training on GPUs requires both multiprocessing and multithreading.

\subsection{Scalability}
We evaluate the scalability of \sys in the EC2 cluster. In this experiment,
we fix the mini-batch size in each trainer and increase the number of trainers
when the number of GPUs increases.

\begin{figure}%
    \centering
    \subfloat[\centering \textsc{ogbn-papers100M}]{{\includegraphics[width=4cm]{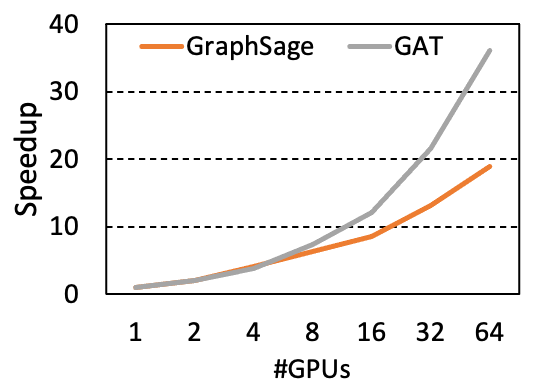} }}%
    \qquad
    \subfloat[\centering \textsc{mag-lsc}]{{\includegraphics[width=3.2cm]{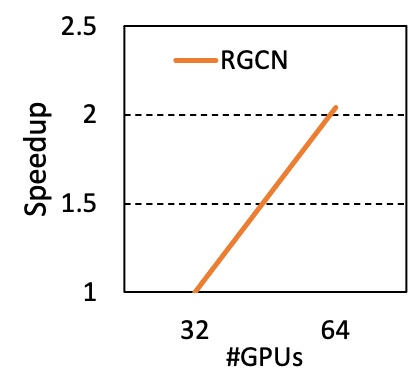} }}%
    \vspace{-3mm}
    \caption{The speedup of \sys with GPUs on large graphs.}%
    \label{fig:scalability}%
\end{figure}

As shown in Figure \ref{fig:scalability}, \sys achieves about 20$\times$ speedup
in GraphSage and 36$\times$ speedup in GAT with 64 GPUs on \textsc{ogbn-papers100M}.
\textsc{mag-lsc} is too large to fit in the CPU memory of one or two g4dn.metal instances.
Its training speed doubles when scaling from
four instances to eight instances. The sub-linear speedup of \sys in GraphSage is
due to CPU saturation caused by mini-batch generation and network saturation caused
by data copy from remote machines. When a GNN model (e.g., GAT) has more
computation, \sys gets better speedup.
In a cluster of 64 GPUs, one epoch takes only 5 seconds for GraphSage and 7 seconds for
GAT on the \textsc{ogbn-papers100M} graph and takes 13 seconds for RGCN
on \textsc{mag-lsc} in a cluster of 64 GPUs.

\subsection{Training convergence}
Each trainer of \sys samples data points from its graph partition, but collectively,
the data points in a global mini-batch are sampled uniformly at random from
the entire training set. This training method is a little similar to ClusterGCN~\cite{clustergcn},
which partitions a graph with METIS and sample partitions to form mini-batches.
We compare \sys with ClusterGCN on OGBN-papers100M. We partition the graph
into 32 partitions for \sys and 16,384 partitions for ClusterGCN.


\begin{figure}[h]
\centering
\includegraphics[scale=0.5]{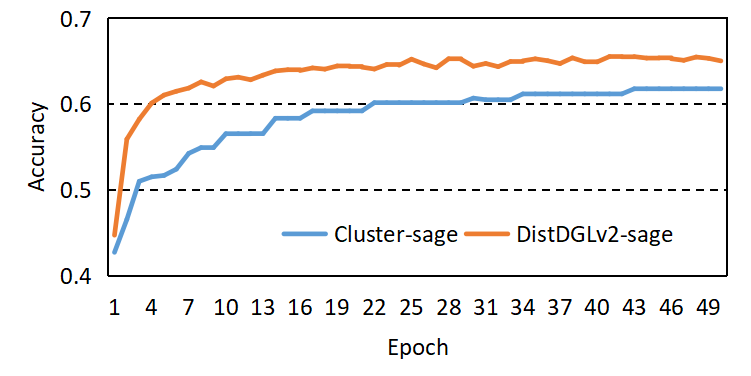}
\vspace{-3mm}
\caption{Convergence of \sys vs. ClusterGCN on \textsc{OGBN-papers100M}.}
\label{fig:converge_papers100M}
\end{figure}

As shown in Figure \ref{fig:converge_papers100M}, we observe slower convergence of
ClusterGCN than \sys and it cannot converge to the same accuracy as \sys.
The main difference between ClusterGCN and \sys is that ClusterGCN drops the edges
that do not belong to the partitions in a mini-batch, while \sys always samples
neighbors uniformly at random.
Thus, \sys estimates neighbor aggregation in an unbiased fashion, while
ClusterGCN's estimation is based by graph partitioning results.
This indicates that we have to sample neighbors across partitions to achieve good model
accuracy.

\begin{table}
    \footnotesize
    \centering
    \caption{Time breakdown of distributed training for different tasks on \textsc{ogbn-papers100M}.}\label{tbl:breakdown}
    \begin{tabular}{@{\hspace{0pt}}lrrrrr@{\hspace{0pt}}}\toprule
         Task & ParMETIS &  Load/save & Load data & Train to \\
                 &          & (partition) & (training) & converge \\
         \midrule
         Vertex classification & 12 min & 23 min & 8 min & 4 min \\
         Link prediction & 12 min & 23 min & 8 min & 305 min \\
         \bottomrule
    \end{tabular}
\end{table}

\subsection{Time breakdown}
In \sys, training a GNN model requires to partition a graph and run a distributed training
job on the partitions. We measure the time of different components in the training pipeline,
including loading and saving data for partitioning, partitioning the graph,
loading partition data for training and finally training
a model to converge. We use ParMETIS~\cite{karypis11parmetis}, a distributed version of METIS,
to partition large graphs. We benchmark ParMETIS on \textsc{ogbn-papers100M}
on a cluster of four r5dn.24xlarge instances and distributed training
jobs on a cluster of g4dn.metal instances.

Table \ref{tbl:breakdown} shows the time breakdown in the training pipeline.
It assumes that graph
partition occurs for every distributed training job. In practice, we partition a graph
for multiple training jobs (e.g., parameter searching and testing different models).
Even in this setting, graph partitioning is not the most time-consuming component
in the training pipeline. It takes only 12 minutes to partition \textsc{ogbn-papers100M}
into 512 partitions. In comparison, data loading and saving takes much more time.
For vertex classification, the training time is short because
\textsc{ogbn-papers100M} has a very small training set (1\% of vertices are in the training
set). It is likely to get a large dataset with more labeled vertices.
For link prediction, we may use all edges to train a model, which leads to a training set
with billions of data points. Training a model for link prediction takes multiple hours
even with one epoch.

\subsection{Ablation Study} \label{sec:ablation}
\sys deploys many optimizations. In this section, we study the effectiveness
of the main optimizations introduced by \sys, excluding the ones introduced in DistDGL:
1) 2-level partition that splits the graph for the levels of machines and GPUs;
2) asynchronous pipeline that performs every operation in mini-batch generation
asynchronously to overlap CPU and GPU computation and network I/O;
3) hybrid CPU/GPU sampling that moves some mini-batch sampling computation to GPUs.
We study these optimizations by adding one optimization
after another until we add all optimizations. The last one basically includes
all optimizations in the study.
The study uses the optimization of METIS partitioning as the baseline because
the benefit of METIS partitioning has been demonstrated by DistDGL~\cite{distdgl}.
We use a cluster of four g4dn.metal instances to run the experiments.

\begin{figure}
\centering
\includegraphics[width=0.8\linewidth]{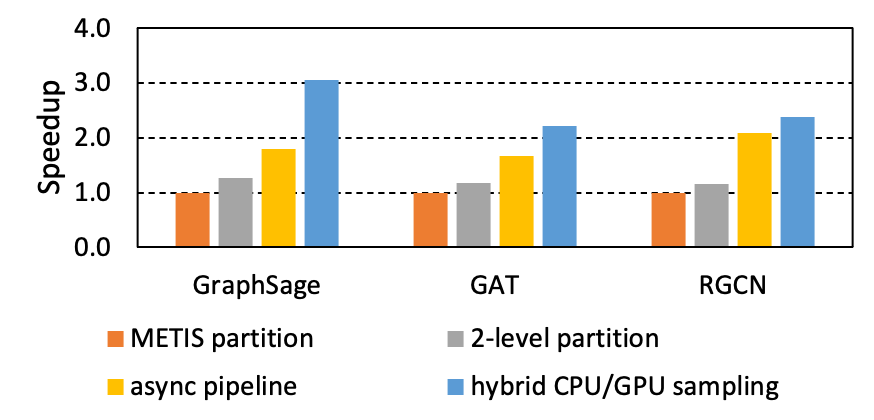}
\vspace{-3mm}
\caption{The effectiveness of various techniques for GraphSage and GAT on
\textsc{ogbn-product} and for RGCN on \textsc{ogbn-mag}.}
\label{fig:opt}
\end{figure}

Figure \ref{fig:opt} shows each optimization has impact in performance and we get
overall $3\times$ speedup for GraphSage and over $2\times$ speedup for GAT and RGCN
on top of METIS partitioning. Even though this cluster already has 100Gbps network,
2-level partitioning gets about $20$\% speedup
because confining the training vertices in a smaller partition
leads to better locality and a smaller number of neighbor vertices
in a mini-batch. Asynchronous sampling pipeline gets significant boost because
it overlaps the CPU and GPU
computation and network I/O to hide network latency, PCIe data transfer
and CPU data copy. Hybrid CPU/GPU sampling is another effective optimization.
This indicates that moving more computation to GPU is beneficial to speed up training.
\section{Conclusion}
We develop \sys for distributed GNN training in a GPU cluster. The hybrid CPU/GPU
training allows to scale to very large graphs. We show that distributed hybrid CPU/GPU
training can get speedup by a factor of $3-15$ over distributed CPU training on a graph
with hundreds of millions of vertices.
\sys adopts many optimizations to make GNN training more efficient in
a cluster of GPUs. We show that only using METIS partitioning is insufficient to achieve
good training speed for
distributed hybrid CPU/GPU training.
By deploying an asynchronous pipeline for generating mini-batches, we can effectively
hide the latency of data communication and overlap
CPU and GPU computation. Because asynchronous mini-batch sampling only applies to
immutable data, it does not affect model convergence. By having
all optimizations, \sys gets $2-3\times$ speedup over DistDGL and $18\times$ speedup
over Euler on GPUs. Currently, the heterogeneous graph support in \sys has been
released as part of DGL 0.7 and have been used in production. We plan to release
the asynchronous mini-batch sampling pipeline in DGL's following release.
\bibliographystyle{plain}
\bibliography{distdgl}

\clearpage
\appendix
\section{Appendix}

\subsection{Hyperparameters and software}
We perform hyperparameter search and select a set of hyperparameters that
achieve good model accuracy on these datasets.
For GraphSage and GAT in vertex classification, we use three GNN layers and
the hidden size of 256; the fanout of each layer is 15, 10 and 5.
GAT uses 2 attention heads.
RGCN uses two layers with the hidden size of 1024 and the sampling fanout is 15 and 25.
We use the batch size of 1000 per trainer for GraphSage and 500 for GAT and RGCN
\footnote{GAT and RGCN run out of memory with the batch size of 1000.}.
For link prediction, we run two GraphSage layers to generate embeddings and
the sampling fanout is 25 and 15; the remaining configurations are the same.
We use a cluster of eight AWS EC2 g4dn.metal instances (96 vCPU, 384GB RAM, 8 T4 GPUs each)
for GPU experiments and a cluster of four AWS EC2 r5d.24xlarge instances (96 vCPU, 768GB RAM)
for CPU experiments and data preprocessing. Both clusters connect machines with 100Gbps network.

In all experiments, we use DistDGL in DGL 0.6\footnote{Some of the features in \sys have
been implemented in DGL0.7 and newer releases} and Pytorch 1.8. \sys is implemented based on
DGL 0.6.
For Euler experiments, we use Euler v2.0 and TensorFlow 1.12.

\end{document}